\documentclass[a4paper,fleqn,usenatbib]{mnras}

\usepackage{color}
\usepackage{hyperref}
\usepackage{graphicx}
\usepackage{amssymb}
\usepackage{amsmath}
\usepackage{bm}
\usepackage{url}
\usepackage{xparse}
\usepackage{xspace}
\xspaceaddexceptions{]\}}

\newcommand{\onefig}[3]{%
  \begin{figure}%
    \centerline{\resizebox{\hsize}{!}{\includegraphics*{#1}}}%
    \caption{#3}\label{#2}%
  \end{figure}%
}
\newcommand{\twofig}[4]{%
  \begin{figure*}%
    \centerline{\resizebox{\hsize}{!}{\includegraphics*{#1} \,%
        \includegraphics*{#2}}}%
    \caption{#4}\label{#3}%
  \end{figure*}%
}
\newcommand{\sect}[1]{Sect.~\ref{#1}\xspace}

\newcommand{\app}[1]{Appendix~\ref{#1}\xspace}
\newcommand{\fig}[1]{Fig.~\ref{#1}\xspace}
\newcommand{\eq}[1]{Eq.~(\ref{#1})\xspace}
\DeclareDocumentCommand{\eqs}{m m m o o}{%
  \IfNoValueTF {#4} {%
    Eqs.~(\ref{#1}){\xspace #2} (\ref{#3})%
  }{%
    Eqs.~(\ref{#1}){\xspace #2} (\ref{#3}){\xspace #4} (\ref{#5})%
  }%
}

\newcommand{\lmax}{\ensuremath{\ell_{\mathrm{max}}}\xspace}

\newcommand{\fsky}{\ensuremath{f_{\mathrm{sky}}}\xspace}
\DeclareDocumentCommand{\alm}{O{} O{} O{a}}{\ensuremath{{#3}_{\ell{#1} m{#2}}}\xspace}
\newcommand{\eps}{\ensuremath{\epsilon}\xspace}
\newcommand{\eeps}{\hat{\epsilon}}
\newcommand{\eepsv}{\bm{\hat{\epsilon}}}
\newcommand{\cl}[3][]{C_{\ell{#1}}^{#2 \times #3}}
\newcommand{\clvinv}[2]{\bm{C}_{\ell}^{#1 \times #2 \, -1}}
\newcommand{\ecl}[2]{\widehat{C}_{\ell}^{#1 \times #2}}

\newcommand{\eclv}[2]{\bm{\widehat{C}}_{\ell}^{#1 \times #2}}
\newcommand{\eclvt}[2]{\bm{\widehat{C}}_{\ell}^{#1 \times #2 \, \dagger}}
\newcommand{\eclvinv}[2]{\bm{\widehat{C}}_{\ell}^{#1 \times #2 \, -1}}
\newcommand{\eclts}{\widehat{C}_{\ell}^{\mathrm{TS}}}
\newcommand{\eclbmp}{\widehat{C}_{\ell}^{\mathrm{BMP}}}
\newcommand{\eclemp}{\widehat{C}_{\ell}^{\mathrm{EMP}}}
\newcommand{\wt}[3][]{w^{#2 \times #3}({\theta{#1}})}
\newcommand{\wtvinv}[2]{\bm{w}(\theta)^{#1 \times #2 \, -1}}

\newcommand{\ewtv}[2]{\bm{\widehat{w}}(\theta)^{#1 \times #2}}
\newcommand{\ewtvt}[2]{\bm{\widehat{w}}(\theta)^{#1 \times #2 \, \dagger}}
\newcommand{\ewtvinv}[2]{\bm{\widehat{w}}(\theta)^{#1 \times #2 \, -1}}
\newcommand{\ewtts}{\widehat{w}(\theta)^{\mathrm{TS}}}
\newcommand{\data}{\bm{d}}
\newcommand{\cv}{\bm{C}}
\newcommand{\tcv}{\bm{\widetilde{C}}}
\newcommand{\dlv}{\bm{D}_{\ell}}
\newcommand{\dlpv}{\bm{D}_{\ell^{\prime}}}
\newcommand{\elv}{\bm{E}_{\ell}}
\newcommand{\elpv}{\bm{E}_{\ell^{\prime}}}
\newcommand{\telv}{\bm{\widetilde{E}}_{\ell}}
\newcommand{\telpv}{\bm{\widetilde{E}}_{\ell^{\prime}}}
\newcommand{\bn}{\bm{n}}
\newcommand{\llpo}[1][]{\ensuremath{(2{\ell{#1}} + 1)}\xspace}

\newcommand{\wigner}[6]{\ensuremath{\begin{pmatrix} #1 & #2 & #3 \\ #4 & #5 & #6 \end{pmatrix}}}
\newcommand{\equ}[2][]{\begin{equation}\label{#1}#2\end{equation}}
\newcommand{\eqa}[2][]{\begin{align}\label{#1}#2\end{align}}
\newcommand{\eqm}[2][]{\begin{multline}\label{#1}#2\end{multline}}
\newcommand{\nn}{\nonumber\\}
\newcommand{\aprop}{\mathrel{\vcenter{\offinterlineskip\halign{\hfil$##$\cr\propto\cr\noalign{\kern2pt}\sim\cr\noalign{\kern-2pt}}}}}

\newenvironment{referee}{\bf}{}
\newcommand{\bref}{\begin{referee}}
\newcommand{\eref}{\end{referee}}

\title{Unbiased methods for removing systematics from galaxy clustering measurements}

\author[Elsner et al.]{Franz Elsner,$^{1}$\thanks{E-mail: f.elsner@ucl.ac.uk}
  Boris Leistedt,$^{1, 2}$
  and
  Hiranya V.~Peiris,$^{1}$
\\
$^1$Department of Physics and Astronomy, University College London,
London WC1E 6BT, U.K.\\
$^2$Center for Cosmology and Particle Physics, Department of Physics,
New York University, New York, NY 10003, USA
}

\date{Accepted \dots. Received \dots; in original form \dots}

\pubyear{2015}

\begin{document}
\label{firstpage}
\pagerange{\pageref{firstpage}--\pageref{lastpage}}
\maketitle

\begin{abstract}
  Measuring the angular clustering of galaxies as a function of
  redshift is a powerful method for extracting information from the
  three-dimensional galaxy distribution. The precision of such
  measurements will dramatically increase with ongoing and future
  wide-field galaxy surveys. However, these are also increasingly
  sensitive to observational and astrophysical contaminants. Here, we
  study the statistical properties of three methods proposed for controlling
   such systematics -- template subtraction, basic mode projection, and
  extended mode projection -- all of which make use of externally
  supplied template maps, designed to characterise and capture the
  spatial variations of potential systematic effects. Based on a
  detailed mathematical analysis, and in agreement with simulations,
  we find that the template subtraction method in its original
  formulation returns biased estimates of the galaxy angular
  clustering. We derive closed-form expressions that should be used to
  correct results for this shortcoming. Turning to the basic mode
  projection algorithm, we prove it to be free of any bias, whereas we
  conclude that results computed with extended mode projection are
  biased. Within a simplified setup, we derive analytical expressions
  for the bias and discuss the options for correcting it in more
  realistic configurations. Common to all three methods is an
  increased estimator variance induced by the cleaning process, albeit
  at different levels. These results enable unbiased high-precision
  clustering measurements in the presence of spatially-varying
  systematics, an essential step towards realising the full potential
  of current and planned galaxy surveys.
\end{abstract}


\section{Introduction}
\label{sec:intro}

Over the last decades, cosmological galaxy surveys collecting
statistically representative samples of galaxies over a wide sky area
have been become legion \citep[e.g.,][]{1983ApJS...52...89H,
  1998AJ....115.1693C, 2000AJ....120.1579Y, 2004MNRAS.355..747J,
  2005MNRAS.362..505C, 2006AAONw.110....3D, 2006A&A...457..841I,
  2006AJ....131.1163S, 2010SPIE.7733E..0EK, 2012arXiv1211.0310L,
  2013ExA....35...25D, 2013AAS...22133501F, 2013Msngr.154...35M,
  2015hsa8.conf..148B}. An important method for extracting and
characterising galaxy clustering information is the computation of the
two-point correlation function on the sphere, the angular correlation
function (or its Fourier transform, the angular power spectrum). It
has proved invaluable as a powerful interface to confront theoretical
cosmological models with observational data
\citep[e.g.,][]{1969PASJ...21..221T, 1974ApJS...28...19P,
  1996MNRAS.283..709H, 2002MNRAS.329L..37B, 2002ApJ...571..172Z,
  2004ApJ...606..702T, 2005ApJ...633..560E, 2007MNRAS.378..852P,
  2010MNRAS.401.2148P, 2010MNRAS.404...60R, 2011MNRAS.416.3017B,
  2011ApJ...728...46W, 2013A&A...552A..96B, 2015arXiv150705360C}.

With decreasing statistical error bars that result from a steady
increase in volume probed by current and future surveys, a proper
control of systematic effects, capable of introducing spurious
signals, is becoming more and more challenging. Among others,
contaminants may be the result of inherent survey characteristics
(e.g., survey depth, seeing, or airmass), the details of data
gathering and processing (e.g., imprinted by the image calibration
procedure), or astrophysical foregrounds (e.g., dust extinction), most
of which are spatially varying over the survey footprint. As a result,
comprehensive template libraries of maps describing the variation of
survey properties over the sky have become a standard data product in
state-of-the-art galaxy surveys \citep{2011MNRAS.417.1350R,
  2012MNRAS.424..564R, 2014MNRAS.444....2L, 2015arXiv150705647L}. The
maps can then be used in a science analysis to either verify the
robustness of results, or to actively correct for the impact of
systematics.

Several methods have been proposed in literature to use systematics
templates to study or reduce the contamination of galaxy angular
clustering measurements by signals of non-cosmological origin (e.g.,
\citealp{2002ApJ...579...48S, 2010MNRAS.406..803B,
  2014MNRAS.445....2V}, see also \citealp{2013MNRAS.432.2945H,
  2015arXiv150904290M}). In the template subtraction approach
introduced in \citet{2012ApJ...761...14H} (also applied in
\citealt{2011MNRAS.417.1350R}), the level of cross-correlation between
systematic template maps and data is used to clean angular clustering
estimates of contaminants. An alternative technique, basic mode
projection, excludes specific spatial patterns described through a set
of templates by assigning infinite variance to them
(\citealt{1992ApJ...398..169R}; with applications in, e.g.,
\citealt{1998ApJ...499..555T, 2004PhRvD..69l3003S,
  2009JCAP...09..006S, 2013A&A...549A.111E, 2013MNRAS.435.1857L}). A
variant thereof, extended mode projection, was subsequently introduced
to identify the most important of all available templates prior to the
analysis, to reduce the total number of modes that have to be
marginalised over \citep{2014MNRAS.444....2L, 2014PhRvL.113v1301L}.

While many of the proposed methods seem adequate in reducing the
impact of systematics, some of them were realized to have a
detrimental effect on the galaxy clustering signal (see, e.g., the
discussion in the Appendix of \citealt{2012MNRAS.424..564R}). In this
paper, we concentrate on the three systematics mitigation methods
mentioned above and study if and in what way the cleaning procedure
affects the statistical properties of angular clustering estimates. In
particular, we investigate whether the results represent unbiased
estimates of the signal properties, and assess if the application of
cleaning procedures introduces additional variance to the measurement.

The paper is organised as follows. In \sect{sec:tpl_subtraction}, we
provide a detailed discussion of the statistical properties of power
spectrum estimates cleaned using the template subtraction method. We
then turn our focus to basic and extended mode projection in
\sect{sec:mode_projection}, and contrast results obtained in harmonic
space and real space (\sect{sec:acf}). Finally, we summarise our
findings in \sect{sec:conclusions}.

\section{Template subtraction}
\label{sec:tpl_subtraction}

In this section, we discuss the properties of a method proposed by
\citet{2012ApJ...761...14H} to account for systematic effects that may
induce spurious signals. The general idea behind this approach is to
use a set of externally supplied templates to be subtracted from the
data with optimally chosen weights, which in turn are estimated from a
cross-correlation of the template and data. In favour of a transparent
discussion, we will first restrict ourselves to the cleaning of a full
sky data set with a single template here and later generalise our
results to multiple templates on the cut sky
(\sect{sec:tpl_subtraction_bias_cut} and \app{app:tpl_sub}).

\subsection{Analytical bias calculation, full sky}
\label{sec:tpl_subtraction_bias_full}

We first analyse the statistical properties of the proposed estimator.
For a single contaminant that contributes to the observed signal with
unknown amplitude \eps, we can construct a linear data model as a
first order Taylor expansion in the template $f$,
\equ[eq:data_model_single_tpl]{
  d = s + \eps f \, ,
}
where $s$ is the signal to be inferred from the data vector $d$ of the
experiment. It is further assumed that signal and template are
uncorrelated, i.e., their cross-covariance vanishes.

Implicitly assuming a non-vanishing template power spectrum
$\cl{f}{f}$, in \citet{2012ApJ...761...14H}, the authors derive an
estimator for the template cleaned signal power spectrum $\cl{s}{s}$,
\equ[eq:ho_estimator_definition]{
  \eclts = \ecl{d}{d} - \eeps^2 \ecl{f}{f} \, ,
}
where
\equ[eq:ho_cleaning_coefficient]{
  \eeps = \ecl{d}{f} / \ecl{f}{f} \, .
}
For consistency with the proposed method in its original formulation,
we consider $\eeps$ to be a function of the multipole moment $\ell$ in
what follows. We hence obtain
\eqa{
  \eclts &= \ecl{d}{d} - \left( \frac{\ecl{d}{f}}{\ecl{f}{f}} \right)^2
  \ecl{f}{f} \nn &= \ecl{s}{s} - \frac{\left( \ecl{s}{f}
    \right)^2}{\ecl{f}{f}} \, .
}

It is then possible to check if the estimator is unbiased by
calculating the ensemble average of all signal realisations,
\equ{
  \left \langle \eclts \right \rangle = \cl{s}{s} -
  \frac{1}{\cl{f}{f}} \left \langle \left( \ecl{s}{f} \right)^2 \right
  \rangle \, .
}
While it may be reasonable to assume that chance correlations between
signal and template vanish on average, the same is not true for the
square of this product,
\eqa[eq:ho_estimator_chance_square_full]{
  \left \langle \left( \ecl{s}{f} \right)^2 \right \rangle &= \left
  \langle \frac{1}{2\ell + 1} \sum_{m} \alm^{s} \alm^{f \, \ast} \,
  \frac{1}{2\ell + 1} \sum_{m^{\prime}} \alm[][^{\prime}]^{s}
  \alm[][^{\prime}]^{f \, \ast} \right \rangle \nn &= \frac{1}{2\ell +
    1} \cl{s}{s} \cl{f}{f} \, ,
}
where we made use of the statistical isotropy of the signal, $\left
\langle \alm^{s} \alm[][^{\prime}]^{s \, \ast} \right \rangle =
\cl{s}{s} \delta_{m m^{\prime}}$. We therefore obtain for the ensemble
average
\equ[eq:ho_single_tpl_avg_full]{
  \left \langle \eclts \right \rangle = \cl{s}{s} \left(1 -
  \frac{1}{2\ell + 1} \right) \, ,
}
i.e, the estimator is biased low, with a relative bias of $b_{\ell} =
- 1/\llpo$. It can be shown (see \app{app:tpl_sub_mult_tpls_full})
that the bias scales with the number of independent templates used in
the cleaning process,
\equ[eq:ho_mult_tpl_avg_full]{
  b_{\ell} = - n / \llpo \, .
}

\subsection{Analytical bias calculation, cut sky}
\label{sec:tpl_subtraction_bias_cut}

Finally, we extend our full sky results obtained so far to the more
realistic case where data are available only on a fraction of the
sphere. In the following, we will assume that all power spectra have
been calculated with a pseudo-$C_{\ell}$ power spectrum estimation
code \citep{2002ApJ...567....2H} which allows us to develop our
results self-consistently within a common framework. We note, however,
that the use of other (e.g., maximum likelihood) estimators is
possible but leaves our conclusions unchanged.

On the cut sky, the spherical harmonics lose their orthogonality,
which manifests itself in a coupling between formerly uncorrelated
Fourier modes. Here, we have to modify
\eq{eq:ho_estimator_chance_square_full} to take this effect into
account. Making use of the properties of the coupling kernels, we
obtain for the ensemble averaged, mask deconvolved signal power
spectrum
\eqa[eq:ho_estimator_chance_square_cut]{
  \left \langle \left( \ecl{s}{f} \right)^2 \right \rangle
  &= \left \langle \sum_{\ell_{1}} M_{\ell \ell_{1}}^{-1}
  \frac{1}{2\ell_{1} + 1} \sum_{m_{1}}
  a_{\ell_{1} m_{1}}^{s} a_{\ell_{1} m_{1}}^{f \, \ast} \right. \nn
  &\times \left. \sum_{\ell_{2}} M_{\ell \ell_{2}}^{-1}
  \frac{1}{2\ell_{2} + 1} \sum_{m_{2}} a_{\ell_{2} m_{2}}^{s}
  a_{\ell_{2} m_{2}}^{f \, \ast} \right \rangle \nn
  &= \sum_{\ell_{1} \ell_{2} \ell_{3}} \! \left( M_{\ell \ell_{1}}^{-1}
  \right)^2 \frac{1}{2\ell_{1} + 1} M_{\ell_{1} \ell_{2}} M_{\ell_{1}
    \ell_{3}} \cl[_{2}]{s}{s} \cl[_{3}]{f}{f} ,
}
where $M_{\ell \ell^{\prime}}$ are the coupling matrices (see
\app{app:tpl_sub_single_tpls_cut} for formal definitions and details
of the calculation). On the cut sky, \eq{eq:ho_single_tpl_avg_full}
hence takes the more complicated form
\eqm[eq:ho_single_tpl_avg_cut]{
  \left \langle \eclts \right \rangle = \cl{s}{s} \\
  \times \left( 1 - \frac{\sum_{\ell_{1} \ell_{2} \ell_{3}} \left(
    M_{\ell \ell_{1}}^{-1} \right)^2 \frac{1}{2\ell_{1} + 1}
    M_{\ell_{1} \ell_{2}} M_{\ell_{1} \ell_{3}} \cl[_{2}]{s}{s}
    \cl[_{3}]{f}{f}}{\cl{s}{s} \cl{f}{f}} \right) \, ,
}
which correctly reduces to \eq{eq:ho_single_tpl_avg_full} in the full
sky limit. We can then just read off the multipole-dependent relative
bias,
\equ{
  b_{\ell} = - \frac{\sum_{\ell_{1} \ell_{2} \ell_{3}} \left( M_{\ell
      \ell_{1}}^{-1} \right)^2 \frac{1}{2\ell_{1} + 1} M_{\ell_{1}
      \ell_{2}} M_{\ell_{1} \ell_{3}} \cl[_{2}]{s}{s}
    \cl[_{3}]{f}{f}}{\cl{s}{s} \cl{f}{f}} \, .
}

Lastly, we discuss the most general case considered here, the cleaning
of a data set on the cut sky with multiple templates. Building on the
results derived in \app{app:tpl_sub_mult_tpls_full}, the relative bias
is
\eqa[eq:ho_mult_tpl_bias_cut]{
  b_{\ell} &= - \frac{1}{\cl{s}{s}} \left \langle \eclvt{s}{f}
  \eclvinv{f}{f} \eclv{s}{f} \right \rangle \nn
  &= - \frac{1}{\cl{s}{s}} \sum_{i j} \left( \clvinv{f}{f} \right)_{i j}
  \nn
  &\times \sum_{\ell_{1} \ell_{2} \ell_{3}} \left( M_{\ell
    \ell_{1}}^{-1} \right)^2 \frac{1}{2\ell_{1} + 1} M_{\ell_{1}
    \ell_{2}} M_{\ell_{1} \ell_{3}} \cl[_{2}]{s}{s}
  \cl[_{3}]{f_{i}}{f_{j}} \, .
}

We provide an approximate estimate of the bias on the cut sky that
reflects the scaling of the coupling matrices with the sky fraction,
\equ[eq:ho_mult_tpl_bias_approx]{
  b_{\ell} \sim - n / \left(\fsky^2 \llpo \right) \, ,
}
where $n$ is the number of independent templates used in the
analysis. We stress that the accuracy of
\eq{eq:ho_mult_tpl_bias_approx} depends sensitively on the functional
form of signal-, template-, and mask power spectra, and is only valid
for large to intermediate sky fractions.

\subsection{Verification with simulations}
\label{sec:tpl_subtraction_sims}

We now confirm the results obtained thus far using 1000 simulated
Gaussian random maps drawn from a flat power spectrum. To showcase the
biasing effect of the cleaning procedure, we generated a set of ten
independent random templates to be subtracted from the simulated
maps. After a power spectrum analysis on the full sky using the
template subtraction method, \eq{eq:ho_estimator_definition}, we
computed the relative deviation of the averaged recovered power
spectra with respect to the input. The observed multipole dependent
bias, shown in \fig{fig:tpl_sub_bias_full}, is in good agreement with
the analytical estimate.

We then verified our understanding of the cleaning process on the cut
sky. Using an azimuthally symmetric binary mask that restricts the
power spectrum measurement to latitudes of $-40^{\circ} \le b \le
40^{\circ}$ ($\fsky \approx 65\,\%$), we analysed 1000 simulations
using a pseudo-$C_{\ell}$ estimator. After applying the cleaning
procedure, we again computed the bias of the result. As can be seen in
\fig{fig:tpl_sub_bias_cut}, it has increased substantially compared to
the full sky analysis. Aside from the exact analytical estimate, we
also compare the numerical results to the approximate bias description
given in \eq{eq:ho_mult_tpl_bias_approx}. We obtain good agreement
between simulations and analytical expressions.

\onefig{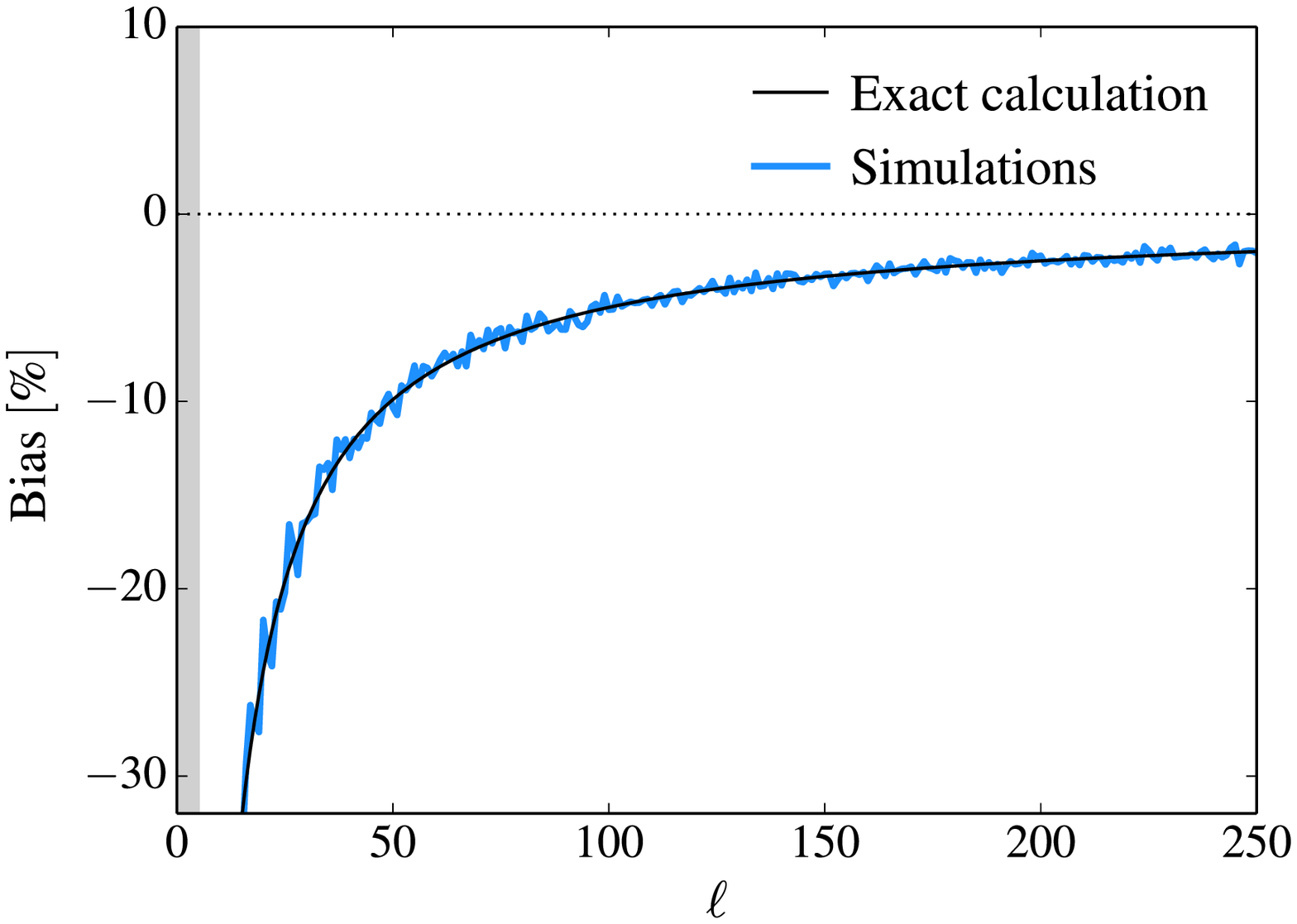}{fig:tpl_sub_bias_full}
{Direct subtraction of systematic templates leads to biased power
  spectrum estimates. Removing contributions from ten templates, we
  show the resulting relative bias of the averaged power spectrum of
  1000 simulations (\emph{blue solid line}) and its analytical
  prediction (\emph{black solid line}). In the multipole region where
  $\ell \le (n-1)/2$, a cleaned power spectrum cannot be constructed
  (\emph{gray region to the left}).}

\onefig{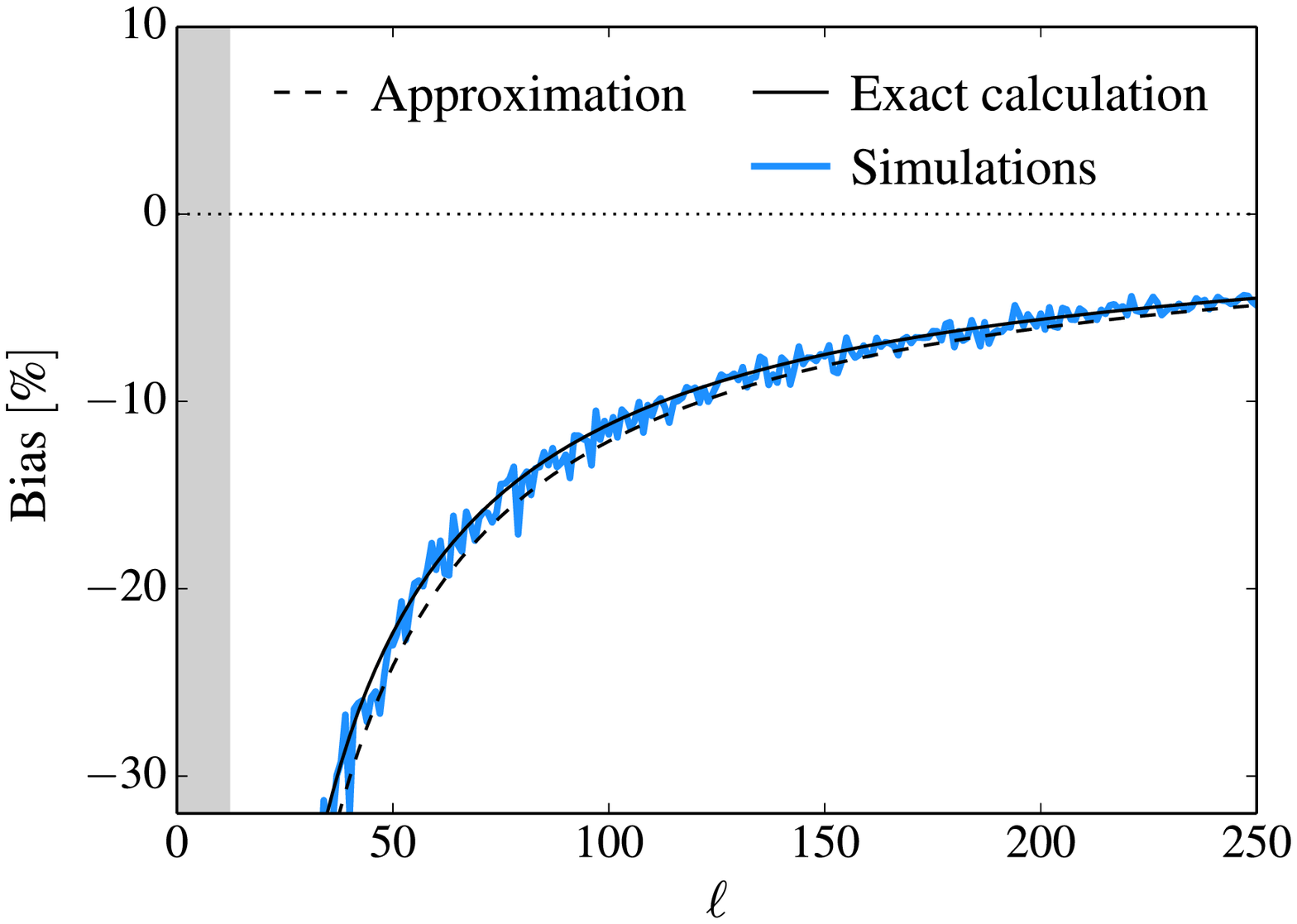}{fig:tpl_sub_bias_cut}
{Same as \fig{fig:tpl_sub_bias_full}, but for partial sky coverage.
  The bias becomes larger for a cut sky analysis. Keeping all other
  simulation parameters the same, we computed pseudo-$C_{\ell}$ power spectra
  on about $65\,\%$ of the sky. The approximate analytical bias
  estimate, obtained from a simple rescaling of the full-sky results,
  is also shown (\emph{black dashed line}).}

\subsection{Discussion}
\label{sec:tpl_subtraction_discussion}

The bias identified in the previous paragraphs can be understood
intuitively. Even if signal and template are uncorrelated on average,
there will still be chance correlations for every individual signal
realisation. By construction, the cleaning procedure will then
minimise the cross-correlation between signal and template, thereby
leading to an over-correction in the ensemble mean. To be more
precise, since the cleaning coefficients $\eeps$ are computed for all
multipole moments individually, at a given $\ell$, the power in one of
the \llpo available Fourier modes will be removed by each template,
giving rise to the simple expression in \eq{eq:ho_mult_tpl_avg_full}.
From a more formal point of view, the problem arises from the fact
that while we derive an estimate of the cleaning coefficient $\eeps$
in \eq{eq:ho_cleaning_coefficient}, we use its square (i.e., a
non-linear transform of it) in the cleaning step
\eq{eq:ho_estimator_definition}. Although we indeed obtain $\left
\langle \eeps \right \rangle = \eps$ on average, we however find
$\left \langle \eeps^2 \right \rangle \ne \eps^2$.

It is interesting to discuss the limiting case where the number of
independent Fourier modes in the data drops below the number of
cleaning templates, $\llpo < n$ for a full-sky data set. For each of
these multipole moments $\ell$, the $n \times n$ matrix constructed
from all possible template auto- and cross-power spectra becomes rank
deficient. As a result, a unique solution for the cleaning
coefficients $\eeps$ no longer exists.\footnote{Of course, it is still
  possible to obtain a (non-unique) solution to the system of
  equations, for example by means of the Moore-Penrose
  pseudo-inverse.} What is more, and this also includes the case where
$2\ell + 1 = n$, the template-subtracted power spectrum estimates
$\ecl{s}{s}$ vanish by construction and a meaningful conclusion about
the cleaned signal amplitude cannot be drawn.

Owing to the aggressive scaling with the sky fraction, $b_{\ell}
\aprop \fsky^{-2}$, on the cut sky the bias can become substantial
even when the data set is cleaned with only a single template.
Fortunately, since we derived closed-form expressions for the bias, we
can naturally propose a correction procedure: to obtain unbiased power
spectrum estimates only requires the multiplication of all
$\ecl{s}{s}$ estimates with the multipole-dependent factor $1 / (1 +
b_{\ell})$, where $b_{\ell}$ is given by one of the expressions in
\eqs{eq:ho_mult_tpl_avg_full}{,}{eq:ho_mult_tpl_bias_cut}[,
  or][eq:ho_mult_tpl_bias_approx]. As discussed above, depending on
the number of templates used, a correction will not be possible for
multipoles below a certain $\ell_{\mathrm{min}}$.

It is also worth noting that the cleaning procedure comes at a
price. Since the effective number of Fourier modes available to
measure $\ecl{s}{s}$ is reduced, the final bias-corrected power
spectrum estimates will suffer from excess variance (i.e., they have
larger error bars). More quantitatively, while a cosmic
variance-limited estimate of a power spectrum computed on the full sky
has variance $\mathrm{Var}(\widehat{C}_{\ell}) = 2 C_{\ell}^2 /
\llpo$, the template cleaning process increases the uncertainty in the
measurement to $\mathrm{Var}(\ecl{s}{s}) = 2 (\cl{s}{s})^2 / (2\ell +
1 - n)$. The impact of the cleaning process on the error bars of
clustering measurements has not been identified and addressed in
previous applications of this method.

While the cleaning coefficient $\eeps$ in
\eq{eq:ho_estimator_definition} is a function of multipole moment in
the original formulation of the algorithm, other authors have assumed
it constant within power spectrum bins
\citep[e.g.,][]{2015arXiv150705551G}, or decided to keep it fixed
entirely \citep[e.g.,][]{2011MNRAS.417.1350R}. Restricting the number
of free parameters in the cleaning procedure will then result in a
reduced bias, since chance correlations are removed only to a lesser
extent. However, these approaches also leave less freedom in case the
systematics templates can only approximately capture the signal
contamination, and may therefore increase the systematics residuals in
the cleaned clustering estimate. Since an analytical calculation of
the bias is no longer possible for these variants of the template
subtraction method, simulations would be required to correct
clustering estimates in practical applications.

Our analytical studies also provide a more principled explanation for
the results of numerical tests conducted in
\citet{2012MNRAS.424..564R} on mock galaxy catalogues to aid the
analysis of \emph{BOSS} data. While the authors compare variations of
the template subtraction method which only allow for a qualitative
comparison, they also find that the cleaning procedure results in a
negative bias of galaxy clustering estimates, in agreement with the
results presented here.

We note in passing that results obtained in this section resemble the
bias identified in so called internal linear combination (ILC) maps
constructed from multi-frequency observations of the cosmic microwave
background radiation \citep[see the discussion in,
  e.g.,][]{2007ApJS..170..288H, 2008PhRvD..78b3003S}.

\section{Mode projection}
\label{sec:mode_projection}

An alternative method for subtracting a set of templates from a
data set was proposed by \citet{1992ApJ...398..169R}. It can be
straightforwardly included into the optimal quadratic power spectrum
estimator for which the computation is based on inverse variance
weighted combinations of the data vector \citep{1997PhRvD..55.5895T}.

\subsection{Basic mode projection}
\label{sec:basic_mp}

We first discuss the basic mode projection approach marginalising over
a single template. The underlying idea of this technique is to modify
the signal covariance matrix to assign infinite variance to modes that
are to be excluded from the analysis. In the following, we verify the
unbiasedness of the derived power spectra.

\subsubsection{Analytical bias calculation}
\label{sec:basic_mp_bias}

Let us first review the basic equations of the optimal quadratic
estimator \citep{1997PhRvD..55.5895T}. To simplify the discussion, we
will consider a full-sky analysis of noiseless data and choose the
spherical harmonic space as basis for our calculations. Retaining the
data model defined in \eq{eq:data_model_single_tpl}, the estimator
derives the power spectrum from a quadratic combination of the data,
\equ{
  \ecl{s}{s} = \sum_{\ell^{\prime}} N_{\ell \ell^{\prime}}^{-1}
  \data^{\dagger} \elpv \data \, ,
}
where $N$ is the estimator normalisation given by the Fisher matrix
$N_{\ell \ell^{\prime}} = 2 F_{\ell \ell^{\prime}}$. Here, the $(\lmax
+ 1)^2 \times (\lmax + 1)^2$ matrices
\equ{
  \elv = \cv^{-1} \frac{\partial \cv}{\partial C_{\ell}} \cv^{-1}
}
are expressed as a function of the covariance matrix
\equ{
  \cv = \sum_{\ell} C_{\ell} \dlv \, ,
}
which takes a particularly simple form since we work in spherical
harmonic space. The matrices $\dlv$ are diagonal and of rank \llpo,
\equ{
  \left( \dlv \right)_{i j} =
  \begin{cases}
    \delta_{i j} & \ell^2 < i \le (\ell + 1)^2 \\
    0 & \mathrm{otherwise}
  \end{cases} \, ,
}
trivially fulfilling the useful relation $\dlv \dlpv = \dlv \,
\delta_{\ell \ell^{\prime}}$.

Mode projection is now included by means of a rank-one update to the
covariance matrix in the equations above, $\tcv = \lim_{\sigma \to
  \infty} \cv + \sigma f f^{\dagger}$. The Sherman-Morrison formula
allows the inverse to be calculated exactly,
\equ{
  \tcv^{-1} = \cv^{-1} - \frac{\cv^{-1} f f^{\dagger}
    \cv^{-1}}{f^{\dagger} \cv^{-1} f} \, .
}

We obtain for the ensemble-averaged, unnormalised signal power spectrum
estimate
\equ[eq:oqe_simple_mp_mean]{
  \left \langle \data^{\dagger} \telv \data \right \rangle =
  \frac{2\ell + 1}{\cl{s}{s}}\left(1 - \frac{\cl{f}{f} /
    \cl{s}{s}}{\sum_{\ell^{\prime}} \llpo[^{\prime}]
    \cl[^\prime]{f}{f} / \cl[^\prime]{s}{s}} \right) \, ,
}
an identity which we derive in \app{app:bmp_mean_var}. The diagonal
elements of the normalisation factor, used to calibrate the estimator,
are
\eqa[eq:oqe_simple_mp_var_diag]{
  \widetilde{N}_{\ell \ell} &= \mathrm{tr} \left( \tcv^{-1} \frac{\partial
    \tcv}{\partial C_{\ell}} \tcv^{-1} \frac{\partial \tcv}{\partial
    C_{\ell}} \right) \nn
  &= \frac{2\ell + 1}{\left( \cl{s}{s} \right)^2}\left(1 - \frac{2
    \cl{f}{f} / \cl{s}{s}}{\sum_{\ell^{\prime}} \llpo[^{\prime}]
    \cl[^\prime]{f}{f} / \cl[^\prime]{s}{s}} \right. \nn
  & \hphantom{= \left. \frac{2\ell + 1}{\left( \cl{s}{s} \right)^2}
    \right)} + \left. \frac{\llpo \left( \cl{f}{f} / \cl{s}{s}
    \right)^2}{\left( \sum_{\ell^{\prime}} \llpo[^{\prime}]
    \cl[^\prime]{f}{f} / \cl[^\prime]{s}{s} \right)^2} \right) \, ,
}
as detailed in \app{app:bmp_mean_var}. Note that although we work on
the full sky, once mode projection is included the Fisher matrix is no
longer diagonal,
\eqa[eq:oqe_simple_mp_var_offd]{
  \widetilde{N}_{\ell \ell^{\prime}} &= \mathrm{tr} \left( \tcv^{-1}
  \frac{\partial \tcv}{\partial C_{\ell}} \tcv^{-1} \frac{\partial
    \tcv}{\partial C_{\ell^{\prime}}} \right) \nn
  &= \frac{2\ell + 1}{\cl{s}{s} \cl[^\prime]{s}{s}} \frac{\cl{f}{f} /
    \cl{s}{s} \llpo[^{\prime}] \cl[^\prime]{f}{f} /
    \cl[^\prime]{s}{s} }{\left( \sum_{\ell^{\prime \prime}}
    \llpo[^{\prime \prime}] \cl[^{\prime \prime}]{f}{f} /
    \cl[^{\prime \prime}]{s}{s} \right)^2}
}
for $\ell \ne \ell^{\prime}$. This newly introduced effect of mode
coupling is in agreement with the interpretation that mode projection
is in fact equivalent to masking. It is then possible to prove that
\eqa{
  \left \langle \eclbmp \right \rangle &= \sum_{\ell^{\prime}}
  \widetilde{N}_{\ell \ell^{\prime}}^{-1} \left \langle \data^{\dagger}
  \telpv \data \right \rangle \nn
  &= \cl{s}{s}
}
(see \app{app:bmp_bias_proof}), i.e., the basic mode projection
algorithm is unbiased.

\subsubsection{Verification with simulations}
\label{sec:basic_mp_sims}

To verify results obtained for the basic mode projection algorithm, we
applied the technique to a set of 1000 simulated maps and templates
with the same properties as previously introduced in
\sect{sec:tpl_subtraction_sims}. We used the identical mask ($\fsky
\approx 65 \%$) and restricted the analysis to relatively low
resolution to accommodate the comparatively high computational
complexity of the optimal quadratic estimator. We show binned
measurements of the relative bias computed from the averaged power
spectrum in \fig{fig:bmp_bias_cut}, where we projected out ten
independent templates. In agreement with the theoretical analysis
presented above, we find no evidence for a bias in power spectrum
measurements with basic mode projection. Comparing the diagonal
elements of the Fisher matrix with and without mode projection
enabled, we also show the increase in error bars of the power spectrum
coefficients induced by the cleaning procedure.

\onefig{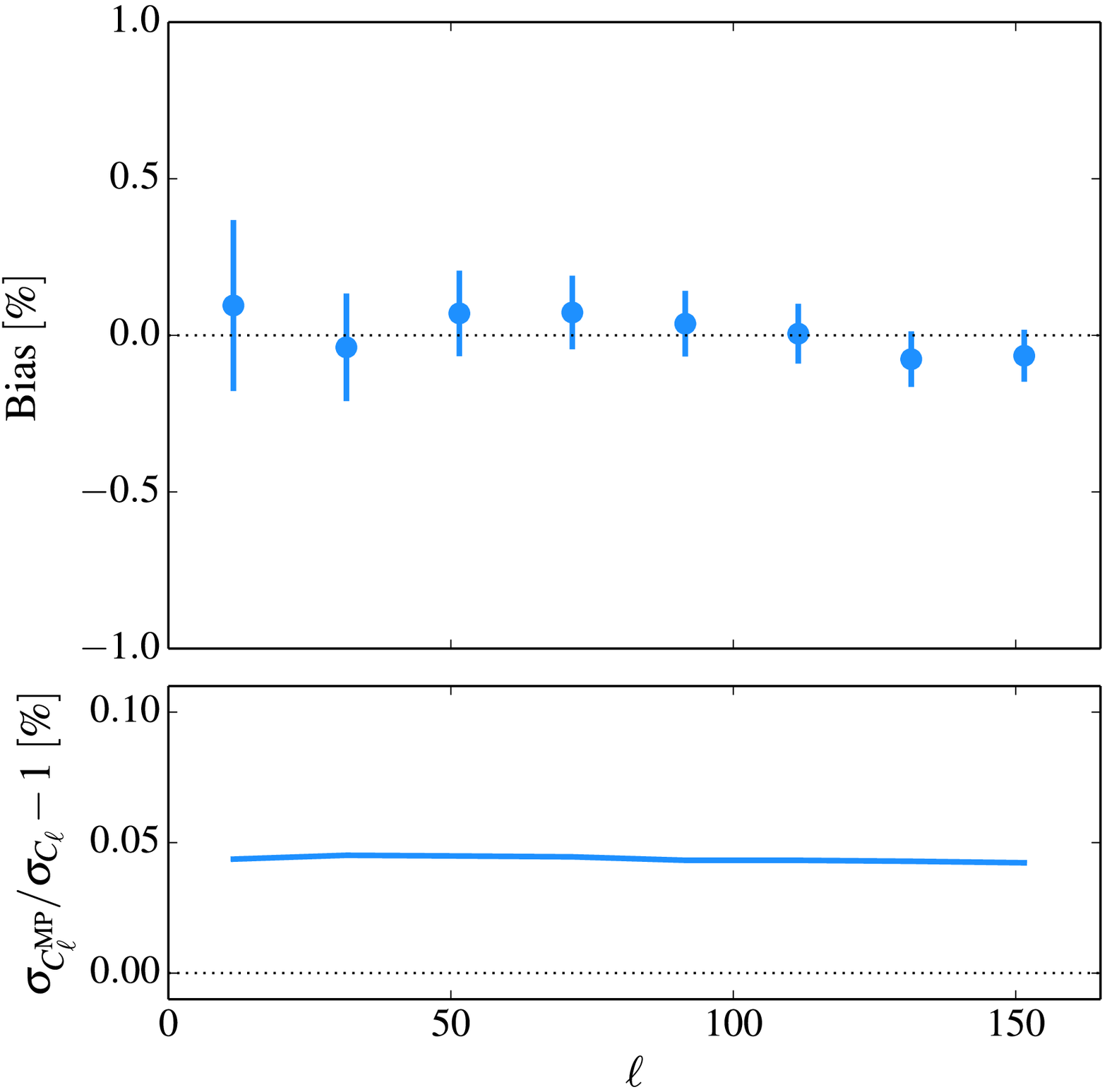}{fig:bmp_bias_cut}
{Basic mode projection leads to unbiased power spectrum
  estimates. \emph{Top panel}: the average power spectrum estimate of
  1000 simulated maps, computed on the cut sky with an optimal
  quadratic estimator with basic mode projection, is consistent with
  the input over the full multipole range (\emph{blue
    circles}). \emph{Bottom panel}: the increase in error bars of the
  power spectrum measurement as a result of mode projection is in the
  sub-percentage regime.}

\subsubsection{Discussion}
\label{sec:basic_mp_discussion}

Unlike the template subtraction approach discussed in
\sect{sec:tpl_subtraction}, basic mode projection correctly accounts
for the reduced variance in power spectrum estimates in
\eq{eq:oqe_simple_mp_mean} with an appropriately rescaled expression
for the estimator normalisation
\eqs{eq:oqe_simple_mp_var_diag}{,}{eq:oqe_simple_mp_var_offd}. Although
we restricted the analytical derivations in this section to the case
of a single template, this conclusion seems to hold for arbitrary
numbers of templates, as indicated by the results of our numerical
studies. Likewise, since the quadratic estimator takes into account
the analysis mask in a mathematically exact way, the discussion
extends to the cut-sky case.

The inverse of the normalisation provides us with direct access to the
estimator variance. In contrast to the template subtraction method in
\sect{sec:tpl_subtraction}, the more complicated mathematical
expressions make a straightforward interpretation difficult. If we
assume, however, a constant ratio of template to signal power spectra,
we can gain some insights by deriving an approximate expression for
the variance of the signal power spectrum estimate. Considering the
case of a full-sky data set with mode projection of a single template,
we obtain
\equ{
  \mathrm{Var}(\ecl{s}{s}) \approx \frac{2 (\cl{s}{s})^2}{2\ell + 1}
  \left( 1 + \frac{1}{(\lmax + 1)^2} \right) \, .
}
Alternatively, we can express the increased variance in terms of a
reduced effective sky fraction of the experiment,
\equ{
  \mathrm{Var}(\ecl{s}{s}) \approx \frac{2 (\cl{s}{s})^2}{(2\ell + 1)
    \, \fsky^{\mathrm{eff}}} \, ,
}
where we have defined $\fsky^{\mathrm{eff}} \approx 1 - 1 / (\lmax +
1)^2 < 1$. Here, the excess variance is approximately given by the
ratio of the single template mode projected to the total number of
Fourier modes present in a data set band-limited at \lmax. We observe
a smaller increase in variance compared to the template subtraction
method, where the cleaning procedure is applied at every multipole
moment independently. Still, some of the discussion in
\sect{sec:tpl_subtraction_discussion} also applies to mode projection:
in case the number of templates is too large compared to the number of
Fourier modes at multipole moment $\ell$, power spectrum estimation is
rendered impossible.

\subsection{Extended mode projection}
\label{sec:extended_mp}

Although basic mode projection is unbiased, including template
marginalisation over a large number, possibly thousands, of
systematics maps will result in a substantial increase in estimator
variance, considerably degrading the predictive power of the data
set. This observation prompted the development of the extended mode
projection algorithm \citep{2014MNRAS.444....2L}, where a smaller
subset of templates to be projected is selected among all available
templates by means of some heuristic criterion prior to the power
spectrum analysis. The idea behind the selection process is to
identify systematic maps that show noticeable correlations with the
data and may therefore be adequate tracers of spurious
signals. Systematic maps that are not or only slightly correlated with
the data, on the other hand, appear to lack relevance for describing
possible contaminants and are therefore excluded from being
marginalised over.

\subsubsection{Analytical bias calculation}
\label{sec:extended_mp_bias}

In the following, we assess the effects of the template selection step
on the statistical properties of the power spectrum estimate. To do
so, we first have to specify a selection criterion. While an
approximate $\chi^2$ estimate computed from the cross-power spectrum
of signal and template was used in the original formulation of the
algorithm \citep{2014MNRAS.444....2L}, here we adopt a simplified
measure to enable a more transparent discussion. We consider a full
sky experiment and a template $f$, containing power only at a single
multipole moment $\alm[][][f] \propto \delta_{\ell \ell^{\prime}}
\delta_{m 0}$. Setting $a_{\ell 0}$ to be the spherical harmonic
coefficient of the data vector corresponding to the template mode, we
adopt a selection criterion based on a predefined threshold $t \ge 0$
such that the extended mode projection algorithm defaults to a
standard power spectrum estimation method if $| a_{\ell 0} | \le t$,
while it otherwise makes use of basic mode projection.

We compute the ensemble average of the signal power spectrum estimate
to check if the estimator is unbiased. For this first test, we assume
that the data is free of any contaminant. In that case,
\eqa[eq:oqe_emp_sys_free_mean]{
  \left \langle \eclemp \right \rangle &= \left \langle \prod_{m \ne 0}
  \int_{-\infty}^{\infty} \mathrm{d} \alm \int_{-\infty}^{-t}
  \mathrm{d} a_{\ell 0} \, P(\{\alm\}) \, \frac{\sum_{m} | \alm
    |^{2}}{2\ell + 1} \right. \nn
  &+ \prod_{m \ne 0} \int_{-\infty}^{\infty} \mathrm{d} \alm
  \int_{-t}^{t} \mathrm{d} a_{\ell 0} \, P(\{\alm\}) \, \frac{\sum_{m
      \ne 0} | \alm |^{2}}{2\ell} \nn
  &+ \left. \prod_{m \ne 0} \int_{-\infty}^{\infty} \mathrm{d} \alm
  \int_{t}^{\infty} \mathrm{d} a_{\ell 0} \, P(\{\alm\}) \,
  \frac{\sum_{m} | \alm |^{2}}{2\ell + 1} \right \rangle \nn
  &= \cl{s}{s} \left( 1 - \sqrt{\frac{2}{\pi \cl{s}{s}}}
  \frac{t}{2\ell + 1} \, \mathrm{e}^{-\frac{t^2}{2 \cl{s}{s}}} \right)
  \, ,
}
for a Gaussian random field \alm. We conclude that for the case
considered here, power spectra estimated with the extended mode
projection algorithm are biased with a relative bias of
\equ[eq:oqe_emp_sys_free_bias]{
  b_{\ell} = - \sqrt{\frac{2}{\pi \cl{s}{s}}} \frac{t}{2\ell + 1} \,
  \mathrm{e}^{-\frac{t^2}{2 \cl{s}{s}}} \, .
}

In the presence of a contaminating signal that can be perfectly
characterised by the template, however, the situation changes. For the
updated data model $d = s + k f$, where $k = \eps \Big /
\sqrt{\cl{s}{s}} \ge 0$ is the relative level of contamination, the
integral bounds in \eq{eq:oqe_emp_sys_free_mean} shift from $\pm t$ to
$\pm t - k$. We then obtain a generalised form of
\eq{eq:oqe_emp_sys_free_bias},
\eqa[eq:oqe_emp_sys_bias]{
  b_{\ell} &= - \sqrt{ \frac{2}{\pi \cl{s}{s}}} \frac{1}{2\llpo}
  \left( t -k + (t + k) \, \mathrm{e}^{\frac{2 t k}{\cl{s}{s}}}
  \right) \, \mathrm{e}^{-\frac{(t + k)^2}{2 \cl{s}{s}}} \nn
  &- \frac{k^2}{2\llpo \cl{s}{s}} \left[ \mathrm{erf} \left( \frac{k -
    t}{\sqrt{2 \cl{s}{s}}} \right) - \mathrm{erf} \left( \frac{k +
    t}{\sqrt{2 \cl{s}{s}}} \right) \right] \, ,
}
where the bias becomes a function of the additional parameter $k$. We
note that the results obtained in
\eqs{eq:oqe_emp_sys_free_bias}{and}{eq:oqe_emp_sys_bias} are only
valid for the selection criterion introduced above.

\subsubsection{Verification with simulations}
\label{sec:extended_mp_sims}

To test our results for correctness, we again compare the analytical
results derived in the previous paragraph to simulations. Since
\eqs{eq:oqe_emp_sys_free_bias}{,}{eq:oqe_emp_sys_bias} are both
functions of the selection threshold $t$, we present results of the
parameter exploration for a fixed multipole moment, $\ell = 5$. To
verify the first case discussed, \eq{eq:oqe_emp_sys_free_bias}, we
simulate $100\,000$ systematics-free data sets. We compute power
spectra using extended mode projection with a single template, where
we chose the values of the threshold parameter from a regularly spaced
grid. To check the second result, \eq{eq:oqe_emp_sys_bias}, we
prepared a new set of simulations, adding a constant contribution of a
contaminant with amplitude $k = 0.5$. In \fig{fig:emp_bias}, the
derived relative bias estimates are plotted as a function of the
selection threshold. Comparing the analytical formula to simulation
results, we find good agreement.

\twofig{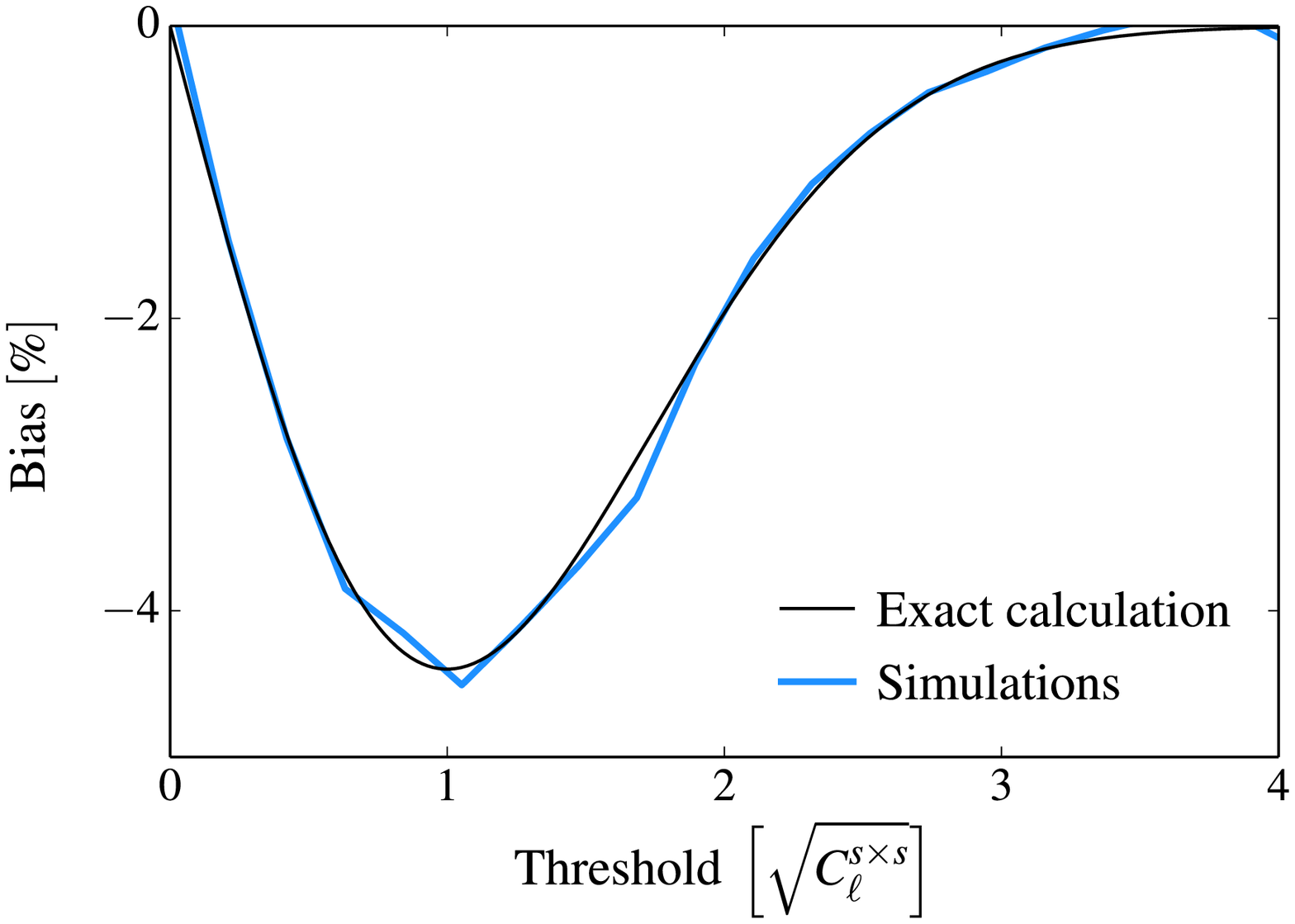}{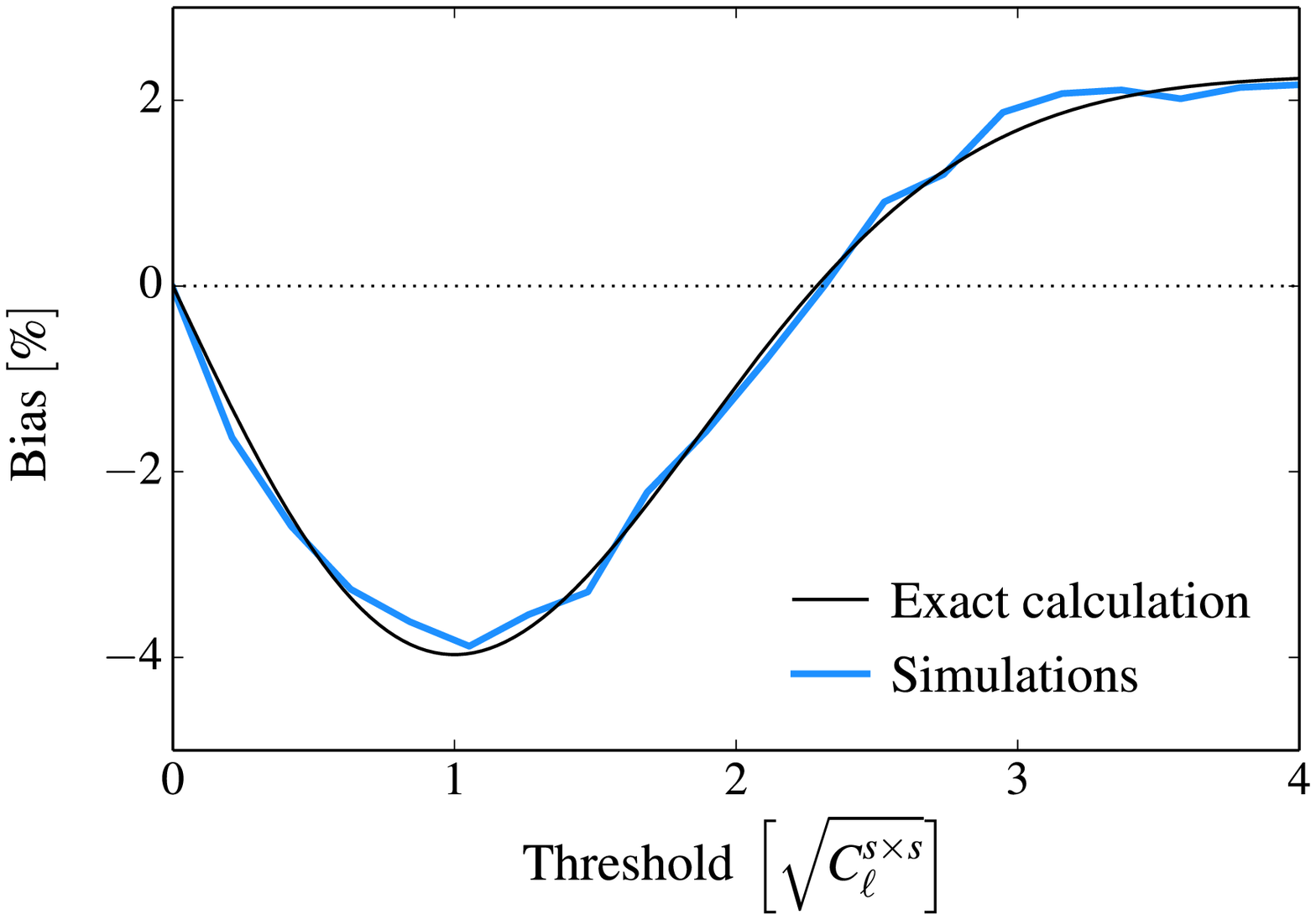}{fig:emp_bias}
{Extended mode projection power spectrum estimates can be biased. For
  a single template analysis, we show the relative bias of the
  algorithm as a function of the template selection threshold from
  simulations (\emph{blue solid lines}) and analytical estimates
  (\emph{black solid lines}) for the multipole moment $\ell = 5$. The
  bias is strictly non-positive in case the data are intrinsically
  free of any contaminants (\emph{left-hand panel}), while it
  otherwise crosses zero (\emph{right-hand panel}).}

\subsubsection{Discussion}
\label{sec:extended_mp_discussion}

Although the extended mode projection algorithm is a very close
derivative of basic mode projection, the two approaches to systematics
mitigation behave qualitatively differently regarding their ensemble
averaged power spectrum estimates. For extended mode projection, we
find in general a non-zero bias whose exact numerical value is
dependent on the template selection criterion adopted in the
analysis. In agreement with results presented in \sect{sec:basic_mp},
the bias vanishes in the limit $t \rightarrow 0$, where extended mode
projection becomes equivalent to basic mode projection. Likewise, if
the data are free of systematics, the bias goes to zero in the limit
$t \rightarrow \infty$, where extended mode projection reduces to
simple power spectrum estimation. The latter result changes, however,
in case there is a contaminant contributing to the observed signal. If
it can be captured by the template, then there exists a non-zero value
of the threshold parameter for which the power spectrum estimates
become unbiased. Unfortunately, to locate this sweet spot would
require knowledge of the actual level of systematics in the data,
which will not be easily available in real-world
applications. Although beyond the scope of this paper, we note that
forward-modelling simulations, attempting to model the full transfer
function of the survey including systematic effects, appear to be
well-suited to provide the additional information needed to debias
signal power spectra \citep{2013A&C.....1...23B, 2013AAS...22134107B,
  2015ApJ...801...73C}.

It is relatively straightforward to identify the source of the
observed bias. Since the template selection process is based on the
actual data realisation, it will inevitably be influenced by chance
correlations between signal and template. As a result, for the
selection criterion adopted here, larger values of the signal
amplitude are more likely to trigger the use of mode projection than
smaller values, leading to an underestimation of the signal variance
on average.

\section{Results for angular correlation function measurements}
\label{sec:acf}

Since we assumed an isotropic signal, results in the previous sections
have been exclusively derived in spherical harmonic space, a basis
where symmetries simplify most of the analytical calculations
considerably. Real space angular correlation functions are widely used
in the field of large scale structure analysis. Their information
content is equivalent; they are related to power spectrum measurements
in a mathematically exact way,
\equ[eq:cl_wtheta]{
  w(\theta) = \sum_{\ell = 0}^{\infty} \frac{2\ell + 1}{4 \pi}
  C_{\ell} P_{\ell} \left( cos(\theta) \right) \, ,
}
where the $P_{\ell}$ are Legendre polynomials of degree $\ell$.

In the following, we discuss the generalisation of our harmonic space
results to real space. Unfortunately, since the most popular angular
correlation function estimator introduced by
\citet{1993ApJ...412...64L} does not use the inverse variance weighted
data vector as basis for the calculation, the mode projection methods
reviewed in \sect{sec:mode_projection} cannot be straightforwardly
extended to real space analyses applying this estimator. We will
therefore only consider the template subtraction method of
\sect{sec:tpl_subtraction}.

\subsection{Analytical bias calculation}
\label{sec:tpl_subtraction_bias_real_space}

To obtain real-space expressions for the bias, we first discuss the
direct transformation of our results from \sect{sec:tpl_subtraction}
to real space, i.e., still assuming that the cleaning procedure itself
was performed in harmonic space. Given an analytical expression for
the multipole dependent relative bias $b_{\ell}$ of the signal power
spectrum, we obtain for the ensemble averaged angular correlation
function
\eqa[eq:wtheta_bias]{
  \left \langle \ewtts \right \rangle &= \wt{s}{s} + \sum_{\ell =
      0}^{\infty} \frac{2\ell + 1}{4 \pi} \left( b_{\ell} \, \cl{s}{s}
    \right) P_{\ell} \left( cos(\theta) \right) \nn
  &= \wt{s}{s} + w^{b}(\theta) \, ,
}
where the real-space bias term $w^{b}(\theta)$ now always implicitly
depends on $\cl{s}{s}$ and cannot be easily expressed as a
multiplicative correction to $\wt{s}{s}$. This is a characteristic
property, typical for the mixing of Fourier modes in angular
correlation function measurements. Given the bias derived in harmonic
space, \eq{eq:ho_mult_tpl_bias_cut} (or its approximation,
\eq{eq:ho_mult_tpl_bias_approx}), it is then possible to obtain
equivalent real space expressions to debias angular correlation
function measurements cleaned with the template subtraction method
using \eq{eq:wtheta_bias}.

In the more realistic case where the full analysis, including the
cleaning step, is performed in real space, the situation grows more
complex. Building on the formalism developed for the harmonic space
analysis in \sect{sec:tpl_subtraction} and \app{app:tpl_sub}, we find
for angular correlation function estimates on the cut sky, cleaned
with multiple templates,
\eqa{
  \left \langle \ewtts \right \rangle &= \wt{s}{s} - \left \langle
  \ewtvt{s}{f} \ewtvinv{f}{f} \ewtv{s}{f} \right \rangle \nn
  &= \wt{s}{s} - \sum_{i j} \left( \wtvinv{f}{f} \right)_{i j}
  \wt{s s}{f_{i} f_{j}} \, ,
}
where we identify the second term on the right-hand side as bias
$w^{b}(\theta)$. If the analysis is performed on the full sky, for
example, we find
\eqm[eq:ho_mult_tpl_bias_full_real]{
  \wt{s s}{f_{i} f_{j}} = \sum_{\ell} \left( \frac{2\ell + 1}{4 \pi}
  P_{\ell} \left( cos(\theta) \right) \right)^2 \\
  \times \frac{1}{2\ell + 1} \cl{s}{s} \cl{f_{i}}{f_{j}} \, .
}
To derive this expression for the bias in real space, we again needed
to make use of the isotropy of the signal in the Fourier domain. As
before, the mode mixing in angular correlation function estimates
prevents us from writing the estimator bias term in
\eq{eq:ho_mult_tpl_bias_full_real} as a signal-independent
multiplicative factor.

\subsection{Verification with simulations}
\label{sec:tpl_subtraction_sims_real_space}

To demonstrate the correctness of our calculation, we computed angular
correlation function estimates of 1000 simulated maps, cleaned with
the template subtraction method in real space. To this end, we
generated band-limited Gaussian maps drawn from a $C_{\ell} \propto
(\ell + 1)^{-1}$ power spectrum, smoothly truncated at \lmax using a
cosine apodisation. We derived correlation function measurements of
the signal on the full sky using ten templates in the cleaning
process, adopting a minimal width for the angular binning set by the
pixel size. In \fig{fig:bmp_bias_cut_real} we show results at several
different values of $\theta$ in comparison with the analytical
estimate obtained from \eq{eq:ho_mult_tpl_bias_full_real}, finding
good agreement. We note that the data points are significantly
correlated across all angular scales.

\onefig{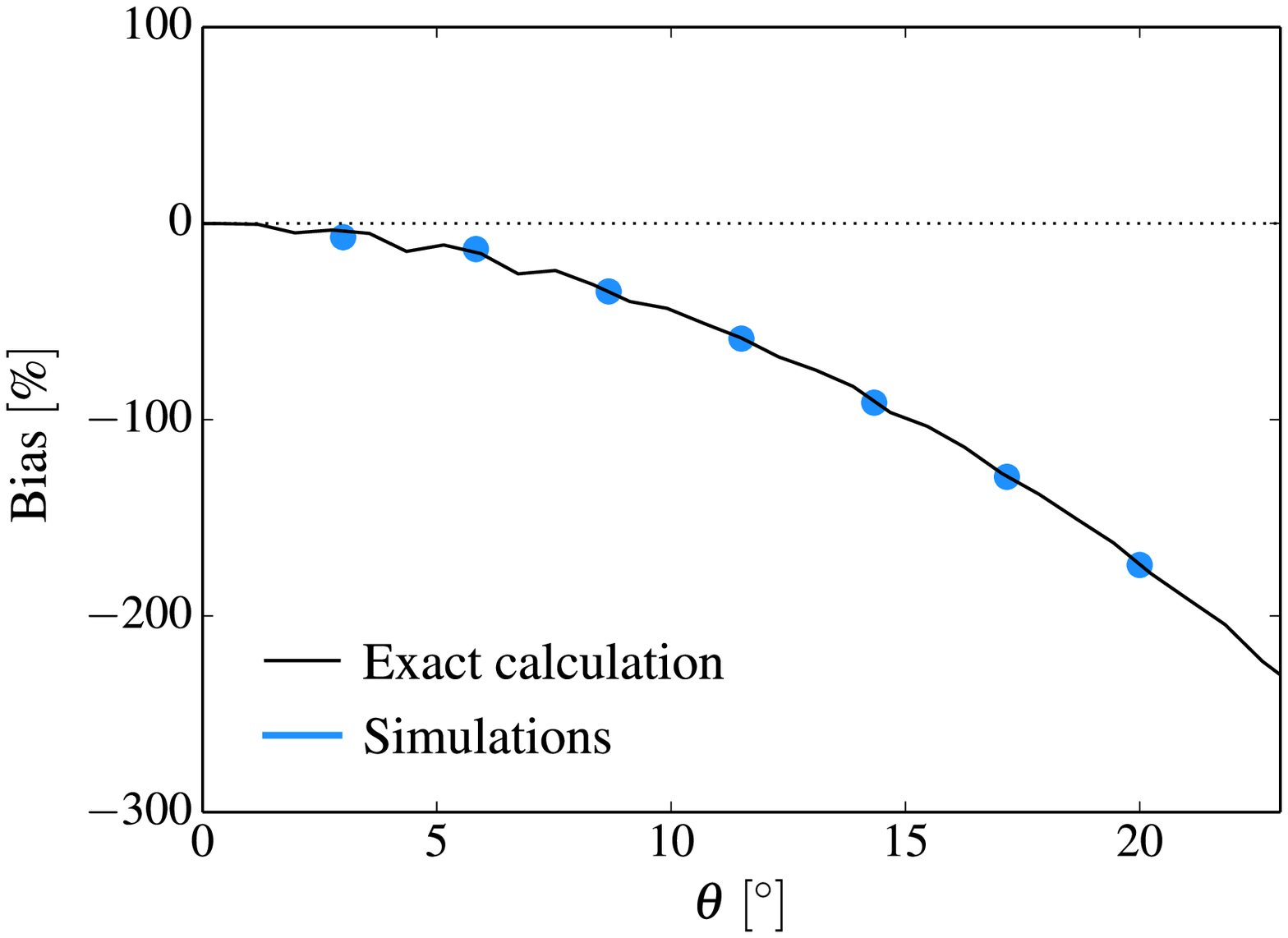}{fig:bmp_bias_cut_real}
{Same as \fig{fig:tpl_sub_bias_full}, but for measurements of the
  angular correlation function in real space. The bias formally
  remains finite at all angular scales as a result of mode coupling.}

\subsection{Discussion}
\label{sec:tpl_subtraction_discussion_real_space}

While the conclusions presented in \sect{sec:tpl_subtraction} remain
qualitatively unchanged, we find that applying template cleaning in
real space leads to somewhat different bias estimates. Unfortunately,
the coupling of Fourier modes in angular correlation function
estimates complicates the analytical bias calculation.

While power spectra are a function of an integer variable $\ell$, the
argument of the angular correlation function is a real number. It is
therefore necessary to adopt some binning scheme for the latter and
average the measurements over angular separations in intervals with
finite width, $\theta_{\mathrm{min}} \le \theta \le
\theta_{\mathrm{max}}$. Since varying the binning will result in a
different bias introduced by the cleaning procedure, we consider the
use of Monte Carlo simulations to correct angular correlation function
estimates cleaned with the template subtraction method to be the
preferred strategy in practical applications. For that reason, we
limited the derivation of analytical results to a single example.

\section{Summary and conclusions}
\label{sec:conclusions}

Measurements of the angular clustering of cosmological data have
become a standard analysis tool in modern cosmology. To mitigate the
impact of systematic effects, capable of introducing spurious signals
of non-cosmological origin, a number of approaches have been proposed
in literature. Concentrating on three popular techniques, template
subtraction \citep{2012ApJ...761...14H}, basic mode projection
\citep{1992ApJ...398..169R}, and extended mode projection
\citep{2014MNRAS.444....2L}, we presented an in-depth discussion of
the effects of the systematics mitigation method on the inferred power
spectrum estimates. Based on a rigorous mathematical analysis, and in
agreement with simulations, we concluded that two out of the three
methods -- template subtraction and extended mode projection -- return
biased estimates of the cleaned signal power spectra. In detail, we
obtain the following results:

For template subtraction, we derived closed-form expressions for the
multipole-dependent bias in the most general case considered here, the
cleaning of data with multiple templates on the cut sky. We explained
its root cause as a consequence of chance correlations between the
signal realisation and the template. Our results then allow debiased
power spectrum estimates to be obtained with this method, or,
equivalently, measurements of the angular correlation function. We
further identified an increase in variance of the estimates and
concluded that for a given number of cleaning templates, the
clustering on specific angular scales can no longer be
measured. Extending the discussion to the template cleaning of angular
correlation function estimates, we obtained consistent results. Since
we found the analytical bias calculation in real space to be more
involved, we proposed to mainly use simulations to obtain the
correction factor needed to debias results.

The analysis of basic mode projection showed that power spectrum
estimates remain unbiased in this framework. We verified analytically
that mode projection increases the estimator variance and introduces
additional coupling between Fourier modes. Owing to the details of the
cleaning process, we found the excess variance to be smaller than for
the template subtraction method.

Lastly, assessing the properties of the extended mode projection
algorithm, we identified the power spectrum estimates to be
biased. Since the basic mode projection algorithm has been proven
bias-free, we showed that it results from the selection process that
was used to decide if a given template should be marginalised over. We
concluded that the bias originates from chance correlations between
the template and the data, which introduce an implicit dependence of
the selection process on the signal amplitude. Although we were able
to obtain analytical expressions that would in principle allow
debiased power spectrum estimates, they depend on the details of the
adopted template selection criterion as well as on the actual level of
contamination, which we presume unknown. We conclude that additional
information, for example provided by forward-modeling simulations of
the data including systematics, is necessary to obtain unbiased signal
estimates.

\section*{Acknowledgements}

We thank our referee, Ashley Ross, for his valuable comments that
helped improving the presentation of our results. We are grateful to
Hans Kristian Eriksen for useful discussions. FE, BL, and HVP were
partially supported by the European Research Council under the
European Community's Seventh Framework Programme (FP7/2007-2013) / ERC
grant agreement no 306478-CosmicDawn. Some of the results in this
paper have been derived using the
HEALPix\footnote{\url{http://healpix.sourceforge.net}}
\citep{2005ApJ...622..759G} package.

\bibliographystyle{mnras}
\bibliography{literature}

\appendix

\section{Template subtraction}
\label{app:tpl_sub}

\subsection{Full sky analysis with multiple templates}
\label{app:tpl_sub_mult_tpls_full}

We now extend our results derived in \sect{sec:tpl_subtraction} for a
single template to an arbitrary number $n$ of linearly independent
templates using the updated data model
\equ[eq:data_model_multi_tpl]{
  d = s + \sum_{i=1}^{n} \eps_{i} f_{i} \, .
}
To this end, we first have to generalise the estimator
\eq{eq:ho_estimator_definition} by transforming it into a matrix
equation,
\equ[eq:ho_multi_template_estimator]{
  \eclts = \ecl{d}{d} - \eepsv^{\dagger} \eclv{f}{f} \eepsv \, ,
}
where $\eepsv^{\dagger} = (\eeps_{1}, \dots, \eeps_{n})$, and
$\left(\eclv{f}{f}\right)_{i j} = \ecl{f_{i}}{f_{j}}$. Contrary to the
single template case, the estimates for $\eepsv$ are now computed by
solving a system of linear equations,
\equ[eq:ho_multi_template_epsilon]{
  \eepsv = \eclvinv{f}{f} \eclv{d}{f} \, ,
}
where the vectors $\left( \eclv{d}{f} \right)_{i} = \ecl{d}{f_i}$. We
find,
\eqa[eq:ho_multi_template_estim_identity]{
  \eclts &= \ecl{d}{d} - \eclvt{d}{f} \eclvinv{f}{f} \eclv{d}{f} \nn
  &= \ecl{s}{s} - \eclvt{s}{f} \eclvinv{f}{f} \eclv{s}{f} \, .
}
To show the equality of the two right-hand side expressions, we first
note that, given \eq{eq:data_model_multi_tpl},
\equ{
  \ecl{d}{d} = \ecl{s}{s} + 2 \sum_{i} \eeps_{i} \ecl{s}{f_{i}} +
  \sum_{i j} \eeps_{i} \eeps_{j} \ecl{f_{i}}{f_{j}} \, .
}
Using $\ecl{d}{f_i} = \ecl{s}{f_i} + \sum_{j} \eeps_{j}
\ecl{f_{i}}{f_{j}}$, we obtain
\eqa{
  \eclvt{d}{f} \eclvinv{f}{f} & \eclv{d}{f} = \eclvt{s}{f}
  \eclvinv{f}{f} \eclv{s}{f} \nn
  &+ 2\sum_{i j q} \left(\eclvinv{f}{f} \right)_{i j} \ecl{f_{j}}{f_{q}}
  \eeps_{q} \ecl{s}{f_{i}} \nn
  &+ \sum_{i j p q} \left(\eclvinv{f}{f} \right)_{i j} \ecl{f_{j}}{f_{q}}
  \eeps_{q} \eeps_{p} \ecl{f_{i}}{f_{p}} \, ,
}
from which \eq{eq:ho_multi_template_estim_identity} follows. Applying
the same procedure as in \eq{eq:ho_estimator_chance_square_full}, we
find for the ensemble averaged signal power spectrum estimate
\eqa{
  \left \langle \eclts \right \rangle &= \cl{s}{s} - \left \langle
  \eclvt{s}{f} \eclvinv{f}{f} \eclv{s}{f} \right \rangle \nn
  &= \cl{s}{s} - \frac{\cl{s}{s}}{2\ell + 1} \sum_{i j}
  \left(\clvinv{f}{f} \right)_{i j} \cl{f_{i}}{f_{j}} \nn
  &= \cl{s}{s} \left( 1 - \frac{n}{2\ell + 1} \right) \, ,
}
i.e., the generalised expression for the relative bias on the full sky
is $b_{\ell} = - n / \llpo$.

\subsection{Cut sky analysis with a single template}
\label{app:tpl_sub_single_tpls_cut}

For completeness, we start by listing the definitions of the
pseudo-$C_{\ell}$ coupling kernels and matrices used in
\sect{sec:tpl_subtraction_bias_cut}, as derived by
\citet{2002ApJ...567....2H}. Setting \alm[][][\overline{a}] to be the
spherical harmonic expansion coefficients of an unmasked map on the
full sky, the effect of an arbitrary real window function $W$ then
results in a related set of coefficients \alm,
\eqa{
  a_{\ell m} &= \sum_{\ell^{\prime} m^{\prime}}
  \alm[^{\prime}][^{\prime}][\overline{a}] \int \mathrm{d} \bn
  Y_{\ell^{\prime} m^{\prime}} W(\bn) Y_{\ell m}^{\ast} \nn
 &= \sum_{\ell^{\prime} m^{\prime}}
  \alm[^{\prime}][^{\prime}][\overline{a}] K_{\ell m \ell^{\prime}
    m^{\prime}} \, .
}
The coupling kernels $K$ can be explicitly expressed in terms of a
product of Gaunt coefficient with \alm[][][w], the spherical harmonic
expansion of the mask,
\eqa[eq:pcl_coupling_kernels]{
  K_{\ell_{1} m_{1} \ell_{2} m_{2}} &= \sum_{\ell_{3} m_{3}}
  \alm[_{3}][_{3}][w] (-1)^{m_{2}} \nn
  &\times \left[ \frac{\llpo[_{1}]\llpo[_{2}]\llpo[_{3}]}{4 \pi}
    \right]^{1/2} \nn
  &\times \wigner{\ell_{1}}{\ell_{2}}{\ell_{3}}{0}{0}{0}
  \wigner{\ell_{1}}{\ell_{2}}{\ell_{3}}{m_{1}}{-m_{2}}{m_{3}} \, ,
}
where the last two factors are Wigner 3j symbols.

The pseudo-$C_{\ell}$ coupling matrices then relate the power spectrum
of the masked coefficients \alm to the full sky coefficients
\alm[][][\overline{a}] such that the pseudo-$C_{\ell}$ estimates
become unbiased, $\left \langle \widehat{C}_{\ell} \right \rangle =
\sum_{\ell^{\prime}} M_{\ell \ell^{\prime}} \left \langle
\widehat{C}_{\ell^{\prime}}^{\mathrm{PCL}} \right \rangle =
C_{\ell}$. Given the orthogonality relations of the Wigner 3j symbols,
they take a particularly simple form,
\equ[eq:pcl_coupling_matrices]{
  M_{\ell_{1} \ell_{2}} = \frac{2\ell_{2} + 1}{4 \pi} \sum_{\ell_{3}}
  \llpo[_{3}] \cl[_{3}]{w}{w}
  \wigner{\ell_{1}}{\ell_{2}}{\ell_{3}}{0}{0}{0}^{2} \, .
}

With the definitions given above, we can now derive the result quoted
in \eq{eq:ho_estimator_chance_square_cut}. For brevity, let $X =
\langle (\ecl{s}{f})^2 \rangle$, then
\eqa{
  X &= \sum_{\substack{\ell_{1} m_{1}
      \\ \ell_{2} m_{2}}} M_{\ell \ell_{1}}^{-1} M_{\ell
    \ell_{2}}^{-1} \frac{1}{2\ell_{1} + 1} \frac{1}{2\ell_{2} + 1} \nn
  &\times \sum_{\ell_{3} m_{3}} \cl[_{3}]{s}{s} K_{\ell_{1}
    m_{1} \ell_{3} m_{3}} K_{\ell_{2} m_{2} \ell_{3} m_{3}}^{\ast} \nn
  &\times \sum_{\ell_{4} m_{4}} \cl[_{4}]{f}{f} K_{\ell_{1}
    m_{1} \ell_{4} m_{4}}^{\ast} K_{\ell_{2} m_{2} \ell_{4} m_{4}} \, .
}
Expanding the coupling kernels using \eq{eq:pcl_coupling_kernels}, we
obtain
\eqa{
  X &= \sum_{\substack{\ell_{1} m_{1} \\ \ell_{2} m_{2}}} M_{\ell
    \ell_{1}}^{-1} M_{\ell \ell_{2}}^{-1} \frac{1}{2\ell_{1} + 1}
  \frac{1}{2\ell_{2} + 1} \nn
  &\times \sum_{\substack{\ell_{3} m_{3} \\ \ell_{4} m_{4}}}
  \cl[_{3}]{s}{s} \cl[_{4}]{w}{w} (-1)^{m_{2} + m_{3} + m_{4}}
  \sqrt{\llpo[_{1}] \llpo[_{2}]} \nn
  &\times \frac{\llpo[_{3}] \llpo[_{4}]}{4 \pi}
  \wigner{\ell_{1}}{\ell_{3}}{\ell_{4}}{0}{0}{0}
  \wigner{\ell_{2}}{\ell_{3}}{\ell_{4}}{0}{0}{0} \nn
  &\times \wigner{\ell_{1}}{\ell_{3}}{\ell_{4}}{m_{1}}{-m_{3}}{m_{4}}
  \wigner{\ell_{2}}{\ell_{3}}{\ell_{4}}{-m_{2}}{m_{3}}{-m_{4}} \nn
  &\times \sum_{\substack{\ell_{5} m_{5} \\ \ell_{6} m_{6}}}
  \cl[_{5}]{f}{f} \cl[_{6}]{w}{w} (-1)^{m_{1} + m_{5} + m_{6}}
  \sqrt{\llpo[_{1}] \llpo[_{2}]} \nn
  &\times \frac{\llpo[_{5}] \llpo[_{6}]}{4 \pi}
  \wigner{\ell_{1}}{\ell_{5}}{\ell_{6}}{0}{0}{0}
  \wigner{\ell_{2}}{\ell_{5}}{\ell_{6}}{0}{0}{0} \nn
  &\times \wigner{\ell_{1}}{\ell_{5}}{\ell_{6}}{-m_{1}}{m_{5}}{-m_{6}}
  \wigner{\ell_{2}}{\ell_{5}}{\ell_{6}}{m_{2}}{-m_{5}}{m_{6}} \, .
}
By first performing the sum over the projective quantum numbers
$m_{3}, m_{4}, m_{5}$, and $m_{6}$, combined with the orthogonality
relations of the Wigner 3j symbols and upon substituting
\eq{eq:pcl_coupling_matrices}, we arrive at the much simplified
expression \eq{eq:ho_estimator_chance_square_cut}.

\section{Basic mode projection}
\label{app:bmp_bias_calculation}

\subsection{Estimator mean and variance}
\label{app:bmp_mean_var}

Here, we prove the identities used in the derivation of
\eqs{eq:oqe_simple_mp_mean}{,}{eq:oqe_simple_mp_var_diag}[,
  and][eq:oqe_simple_mp_var_offd]. To do so, we first obtain
simplified expressions for a series of terms we will make use of in
the process. We find
\equ{
  \mathrm{tr} \left( \dlv f f^{\dagger} \right) = \llpo
  \cl{f}{f} \, ,
}
\equ{
  f^{\dagger} \tcv^{-1} f = \sum_{\ell} \llpo
  \frac{\cl{f}{f}}{\cl{s}{s}} \, ,
}
\eqm{
  \mathrm{tr} \left( \tcv^{-1} f f^{\dagger} \dlv \tcv^{-1} f
  f^{\dagger} \right) \\
  = \llpo \frac{\cl{f}{f}}{\cl{s}{s}} \sum_{\ell^{\prime}}
  \llpo[^{\prime}] \frac{\cl[^\prime]{f}{f}}{\cl[^\prime]{s}{s}}
  \, ,
}
\equ{
  \mathrm{tr} \left( \tcv^{-1} f f^{\dagger} \dlv \tcv^{-1} f
  f^{\dagger} \dlv \right) = \left[ \llpo \frac{\cl{f}{f}}{\cl{s}{s}}
    \right]^{2} \, .
}
Then, from
\eqm{
  \left \langle \data^{\dagger} \telv \data \right \rangle =
  \frac{1}{\cl{s}{s}} \\
  \times \mathrm{tr} \left[ \left(\dlv - \frac{\cv^{-1}
      f f^{\dagger} \dlv}{f^{\dagger} \cv^{-1} f} \right) \left(1 -
    \frac{\cv^{-1} f f^{\dagger}}{f^{\dagger} \cv^{-1} f} \right)
    \right] \, ,
}
\eq{eq:oqe_simple_mp_mean} follows. Likewise, we find for
\eqm{
  \widetilde{N}_{\ell \ell} = \frac{1}{\left( \cl{s}{s} \right)^2} \\
  \times \mathrm{tr} \left[ \left(\dlv - \frac{\cv^{-1} f f^{\dagger}
      \dlv}{f^{\dagger} \cv^{-1} f} \right) \left(\dlv -
    \frac{\cv^{-1} f f^{\dagger} \dlv}{f^{\dagger} \cv^{-1} f} \right)
    \right] \, ,
}
and
\equ{
  \widetilde{N}_{\ell \ell^{\prime}} = \frac{1}{\cl{s}{s}
    \cl[^\prime]{s}{s}} \mathrm{tr} \left(\frac{\cv^{-1} f f^{\dagger}
    \dlv \cv^{-1} f f^{\dagger} \dlpv}{\left(f^{\dagger} \cv^{-1}
    f\right)^2} \right)
}
for $\ell \ne \ell^{\prime}$, and therefore obtain
\eqs{eq:oqe_simple_mp_var_diag}{and}{eq:oqe_simple_mp_var_offd}.

\subsection{Proof of unbiasedness}
\label{app:bmp_bias_proof}

Given our results summarised in
\eqs{eq:oqe_simple_mp_mean}{,}{eq:oqe_simple_mp_var_diag}[,
  and][eq:oqe_simple_mp_var_offd], we now want to show that the
quadratic estimator remains unbiased when mode projection is included,
i.e., $\sum_{\ell^{\prime}} \widetilde{N}_{\ell \ell^{\prime}}^{-1}
\langle \data^{\dagger} \telpv \data \rangle = \cl{s}{s}$. As it turns
out, it is simpler to prove the equivalent expression $\langle
\data^{\dagger} \telv \data \rangle = \sum_{\ell^{\prime}}
\widetilde{N}_{\ell \ell^{\prime}} \cl[^{\prime}]{s}{s}$
instead. Indeed, we find,
\eqa{
  \sum_{\ell^{\prime}} \widetilde{N}_{\ell \ell^{\prime}}
  \cl[^{\prime}]{s}{s}
  &= \frac{2\ell + 1}{\cl{s}{s}} \left( 1 - 2 \frac{\cl{f}{f} /
    \cl{s}{s}}{\sum_{\ell^{\prime}} \llpo[^{\prime}]
    \cl[^\prime]{f}{f} / \cl[^\prime]{s}{s}} \right. \nn
  &+ \frac{\cl{f}{f} / \cl{s}{s}}{\sum_{\ell^{\prime}}
    \llpo[^{\prime}] \cl[^\prime]{f}{f} / \cl[^\prime]{s}{s}} \nn
  &\left. \times \sum_{\ell^{\prime}} \frac{\llpo[^{\prime}]
    \cl[^{\prime}]{f}{f} / \cl[^{\prime}]{s}{s}}{\sum_{\ell^{\prime
        \prime}} \llpo[^{\prime \prime}] \cl[^{\prime \prime}]{f}{f} /
    \cl[^{\prime \prime}]{s}{s}} \right) \nn
  &= \left \langle \data^{\dagger} \telv \data \right \rangle \, .
}

\bsp
\label{lastpage}
\end{document}